\documentclass[fleqn,twoside]{article}
\usepackage{espcrc2}
\usepackage{epsfig}
\newcommand{\bc} {\begin{center}}
\newcommand{\ec} {\end{center}}
\newcommand{\bqa}{\begin{eqnarray}}
\newcommand{\eqa}{\end{eqnarray}}
\newcommand{\nn}{\nonumber}
\newcommand{\msb}{\overline{MS}}

\newcommand{\be}{\beta}
\newcommand{\ga}{\gamma}

\newcommand{\ep}{\epsilon}
\newcommand{\si}{\sigma}
\newcommand{\ka}{\kappa}

\def\lrD{\stackrel{\leftrightarrow}{D}}

\title{
\vskip -100pt{\large
\mbox{} \hfill DESY 03-149\\
\mbox{} \hfill SFB/CPP-03-39\\
\mbox{} \hfill September 2003\\}
\vskip 55pt
Pion parton distribution functions from lattice QCD
\thanks{Talk presented by I. Wetzorke. The work was supported by the
EU IHP Network on Hadron Phenomenolgy from Lattice QCD and by the
DFG under SFB/TR 09-03.}}

\author{I.~Wetzorke\address[NIC]{NIC/DESY Zeuthen,
Platanenallee 6, D-15738 Zeuthen, Germany},
M.~Guagnelli\address[ROMA]{Dipartimento di Fisica, Universit\`a di Roma 
        {\em Tor Vergata} and INFN, Sezione di Roma II,\\
        Via della Ricerca Scientifica 1, I-00133 Rome, Italy},
K.~Jansen\addressmark[NIC], 
F.~Palombi\address{E.~Fermi Research Center, c/o Compendio Viminale, pal.~F,
        I-00184 Rome, Italy},
R.~Petronzio\addressmark[ROMA], A.~Shindler\addressmark[NIC]
(ZeRo collaboration)}
       
\begin{document}

\begin{abstract}
We report on recent results for the pion matrix element of
the twist-2 operator corresponding to the average momentum of
non-singlet quark densities. For the first time finite volume
effects of this matrix element are investigated and come out to be
surprisingly large. We use standard Wilson and non-perturbatively
improved clover actions in order to control better the extrapolation to the
continuum limit. Moreover, we compute, fully non-perturbatively,         
the renormalization group invariant matrix element, which allows a comparison
with experimental results in a broad range of energy scales. 
Finally, we discuss the remaining uncertainties, the extrapolation
to the chiral limit and the quenched approximation.
\vspace*{-2mm}\end{abstract}

\maketitle

\section{INTRODUCTION}\vspace*{-1mm}
Parton distribution functions of hadrons are in the focus
of several theoretical and experimental investigations. 
In recent years lattice QCD calculations have provided estimates
for the lowest moments of parton distribution functions, but so far
most of the results have been obtained at one fixed lattice spacing
only and using perturbative renormalization factors for the
bare operators.

The scope of our current investigation is the pion matrix element of
the twist-2 operator corresponding to the average momentum of
non-singlet quark densities. This operator belongs to two irreducible
representations of the lattice $H(4)$ group. We will concentrate solely
on the matrix element of the operator
\bqa
{\cal O}_{44} (x) &=& \bar\psi(x) \Big[ \gamma_4 \lrD_4 - \frac{1}{3} 
\sum_{k=1}^3 \gamma_k \lrD_k \Big] \frac{\tau^3}{2} \psi(x)\;,\nn
\eqa
since it can be computed at zero external pion momentum and thus provides
a reliable signal. In particular we use Schr\"odinger functional (SF)
boundary conditions \cite{Lues92} for our computation, which allows the
extraction of the matrix element from one large plateau around the middle
of the time extent $T$ of the lattice. 
The correlation function of the matrix element $f_M$ is obtained by inserting
the ${\cal O}_{44}$ operator between two SF pion states ${\cal O}_0$ and
${\cal O}_T$ at the time
boundaries $t=0$ and $T$ and suitable normalization with the
boundary-to-boundary correlator $f_1$ \cite{Alpha,Guag00}
\bqa
f_M(t) &=& a^3\sum_{\rm x,y,z} \langle {\cal O}_0 \; {\cal O}_{44}(x) \;
{\cal O}_T\rangle\nn\\
&\sim& e^{-m_\pi T} \; \langle \pi | {\cal O}_{44} | \pi\rangle
\; \{ 1 + \dots \}\nn\\[2mm]
f_1 &=& - \langle {\cal O}_0 {\cal O}_T\rangle \; \sim \; e^{-m_\pi T} \; .\nn
\eqa
The bare pion matrix element is then defined by
\bqa
\langle x\rangle^{bare}
\equiv \frac{2\ka}{m_\pi} \langle \pi | {\cal O}_{44} | \pi\rangle
= \frac{2\ka}{m_\pi} f_M(t)/f_1 \; ,\nn
\eqa
considering a plateau region around $T/2$ where the relative excited state
contribution $\ep$ is below 0.4\%. The pion mass is obtained from the time
dependence of the improved axial-vector correlator $f^I_A$ following
\cite{Alpha} and requiring $\ep$ smaller than 0.1\%.\vspace*{-1mm}

\section{FINITE VOLUME EFFECTS}\label{finite}\vspace*{-1mm}
Finite volume effects for pion and nucleon matrix elements have not been
studied before, but might influence the precise determination
of moments of parton distribution functions from lattice QCD. We have investigated
the effects of limited lattice extent in detail at $\be$=6.1 ($a$=0.079 fm)
with the non-perturbatively improved clover fermion action and lattice sizes
ranging from $12^3 \times 36$ to $32^3 \times 56$.

\begin{figure}[t]
\bc
\hskip-4mm
\epsfig{file=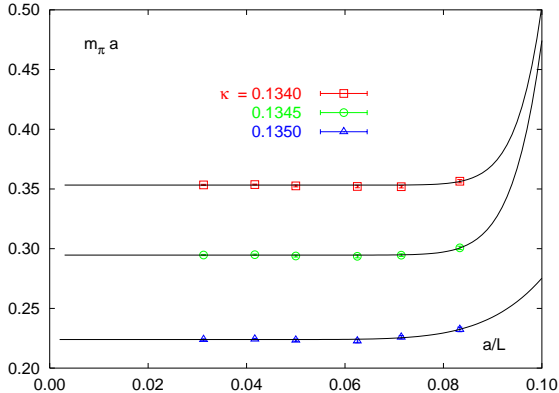,width=78mm}
\vskip-12mm
\ec
\caption{Finite volume effects for $m_\pi a$\vspace*{-5mm}}
\label{f_pi}
\end{figure}

\begin{figure}[t]
\bc
\hskip-4mm
\epsfig{file=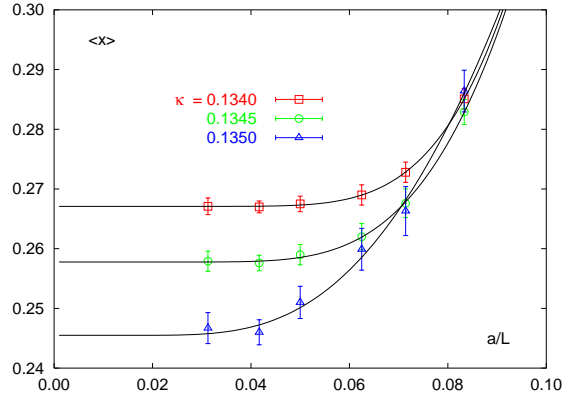,width=78mm}
\vskip-12mm
\ec
\caption{Finite volume effects for $\langle x \rangle$\vspace*{-4.5mm}}
\label{f_X}
\end{figure}

While the pion mass reaches a constant value already for $L \simeq 1.2$
fm and $T \simeq 2.8$ fm (see figure \ref{f_pi}), the bare pion matrix element
in figure \ref{f_X} shows very strong finite volume effects. For the largest
quark mass $L \simeq 1.6$ fm and $T \simeq 3.2$ fm seems to be sufficient,
whereas an even larger spatial extent of 1.9 fm is needed for the smallest
quark mass. The lines in both figures represent exponential fits with the
purely phenomenological ansatz $X(L)\!=\!c_0+c_1/L^{3/2}\exp(-c_2 L)$, where
$X\!\!=\!\!m_\pi a$ or $\langle x \rangle$. Please note that a power-law fit
ansatz would describe the data almost equally well.

These observations lead to the expectation that the finite volume effects
for nucleon matrix elements could be even larger, since the influcence of the
limited lattice size on the nucleon mass has already been observed to be 
much stronger than for the pion mass \cite{Aoki94}. This might be one 
ingredient for the deviation of current lattice QCD estimates of
$\langle x \rangle$ for the nucleon in comparison with experimental data.
\vspace*{-1mm}

\section{CONTINUUM EXTRAPOLATION}\label{cont}
\begin{figure}[t]
\bc
\hskip-4mm
\epsfig{file=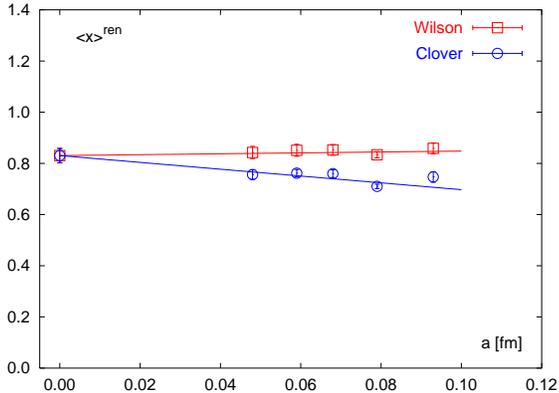,width=78mm}
\vskip-12mm
\ec
\caption{Combined continuum extrapolation for
different discretizations of the fermion action\vspace*{-3mm}}
\label{cont_X}
\end{figure}

\vspace*{-1mm}
In order to improve the continuum extrapolation we use both
the unimproved Wilson and non-perturbatively improved clover
fermion action, which can have different ${\cal O}(a)$ effects
due to the usage of the unimproved ${\cal O}_{44}$ operator,
but should agree in the continuum limit. We have complemented
a previous data set \cite{Guag00} with simulations at $\be$=6.1
$(24^3 \times 42)$ and $\be$=6.45 $(32^3 \times 72)$. The bare matrix
elements were renormalized with the appropriate Z factors computed
non-perturbatively at the same low energy scale $\mu_0$ in the
SF scheme \cite{ZeRo03}.
In figure \ref{cont_X} we show the combined continuum extrapolation
(4 smaller lattice spacings) of the renormalized matrix element
\bqa
\langle x \rangle^{ren}_{SF} (\mu_0) = \lim_{a\to 0} \lim_{m\to 0}
\frac{\langle\pi|{\cal O}_{44}|\pi\rangle(a)}{Z_{SF}(1/\mu_0,a)} \; ,\nn
\eqa
where the chiral extrapolation had been performed linearly in the
quark mass. This yields a value of
$\langle x \rangle^{ren}_{SF} (\mu_0) = 0.870(33)$
for the renormalized pion matrix element in the SF scheme.\vspace*{-1mm}

\section{RGI MATRIX ELEMENT}\label{RGI}\vspace*{-1mm}
The phenomenological analysis of the experimental data
is usually done in the $\msb$ scheme. In order to translate our fully
non-perturbatively computed matrix element
$\langle x \rangle^{ren}_{SF}(\mu_0)$ renormalized in the SF scheme to
the $\msb$ scheme,
we need first to calculate the universal renormalization group invariant
(RGI) matrix element. This is done following the complete non-perturbative
evolution \cite{ZeRo03} in the SF scheme until it is possible to make
(at small volumes) a matching with perturbation theory.
Using the so-called UV-invariant step scaling function it is then possible
to eliminate any reference to the scale $\mu_0$ and obtain finally the
RGI matrix element
\bqa
{\langle x \rangle}^{RGI} = 
{\langle x \rangle}^{ren}_{SF}(\mu_0)\;\si^{UV}_{INV}(\mu_0) = 0.192(11) \; .
\nn
\eqa
The RGI matrix element allows a simple conversion to any desired scheme
(e.g.$\;\msb$ at 2GeV) requiring only the knowledge of the $\be$- and
$\ga$-function up to a certain order in perturbation theory in this scheme
\cite{ZeRo03}.
The resulting number $\langle x \rangle_{\msb} (\mu$=2GeV)=0.265(15)
has decreased slightly compared to the previous lattice
computation \cite{Best97} $\langle x \rangle_{\msb}(\mu$=2.4GeV)=0.273(12),
which has been performed only at one lattice spacing and
with perturbative renormalization factors.\vspace*{-1mm}

\section{LIMITATIONS}\label{limit}\vspace*{-1mm}
Our new result is still higher, but already more precise than the latest NLO
analyses of Drell-Yan and prompt photon $\pi N$ data \cite{SMRS92,Glue99},
which yield a combined experimental value of
$\langle x \rangle_{\msb} (\mu$=2GeV) = 0.21(2).
The quenched approximation is certainly one limitation that has to be explored
in the future.

Another aspect that has been investigated recently is
the chiral extrapolation. It has been shown \cite{Detm03} that a non-linear
fit of the previous data \cite{Best97} leads to 
$\langle x \rangle_{\msb} = 0.24(1)(2)$, which is much more compatible
with the experimental number. The first error is the statistical and the
second describes the variation with the fit parameters. 

\begin{figure}[t]
\bc
\hskip-4mm
\epsfig{file=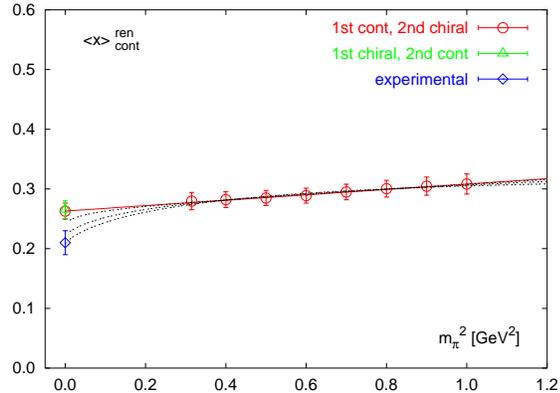,width=78mm}
\vskip-12mm
\ec
\caption{Comparison of chiral expolations:
linear (solid line) and non-linear \cite{Detm03} with
$\Lambda$=0.4, 0.7, 1.0 GeV (dotted lines)\vspace*{-4.5mm}}
\label{chiral_X}
\end{figure}

In figure \ref{chiral_X} we show our results performing first the combined
continuum extrapolation of Wilson and clover data at fixed $m_\pi^2$.
A linear chiral fit yields a consistent result
$\langle x \rangle_{\msb} = 0.263(14)$ with
interchanged limits, while a non-linear fit similar to \cite{Detm03} yields 
$\langle x \rangle_{\msb} = 0.22(1)(2)$ and is thus compatible within error
with the experimental number. Despite these results the correct chiral
extrapolation needs further investigations, especially in the region of
smaller and realistic quark masses.
\vspace*{-1mm}

\end{document}